\documentclass[10pt]{article}
\usepackage{graphicx}

\usepackage{epsfig}
\usepackage{latexsym}
\usepackage{amsmath}

\begin{document}

\title{   
 Coherent measures of the impact of co-authors  in peer review  journals and in proceedings publications 
}

\author{Marcel AUSLOOS 
 \\ 
\\Ê\\ÊRoyal Netherlands Academy of Arts and Sciences\\
Joan Muyskenweg 25, 1096 CJ Amsterdam, The Netherlands    \\Ê\\ÊSchool of Management, University of Leicester, \\ University Road, Leicester  LE1 7RH, UK  \\ \\ GRAPES, r. Belle Jardini\`ere, 483, \\ B-4031 Liege, Wallonia-Brussels Federation\
 }

 \date{\today}
\maketitle
 \vskip 0.5 cm

\begin{abstract}

  This paper focuses on the coauthor effect in different types of publications, usually not equally respected in measuring research impact.   {\it A priori}  unexpected
relationships are found  between the total coauthor  core value, $m_a$,  of a   leading investigator  (LI),   and the related values for their publications in either peer review journals ($j$) or in proceedings ($p$). A surprisingly  linear relationship is found: $  m_a^{(j)} +   0.4\;m_a^{(p)} = m_a^{(jp)}  $.  Furthermore, another relationship is found concerning  the measure of the total number of citations, $A_a$,  i.e. the surface of the  citation size-rank histogram up to $m_a$. Another linear relationship exists : $A_a^{(j)} + 1.36\; A_a^{(p)}  = A_a^{(jp)} $.  These empirical findings coefficients (0.4 and 1.36)  are supported by considerations  based on an empirical power law  found between  the number of joint publications of an author  and the rank  of a coauthor.  Moreover, a simple power law relationship is found between $m_a$ and  the number ($r_M$) of  coauthors of a LI:  $m_a\simeq r_M^{\mu}$; the power law exponent $\mu$ depends on the type ($j$ or $p$) of publications.    These  simple relations, at this time limited to publications in physics,  imply that  coauthors  are a "more positive measure"  of a principal investigator role, in both types of scientific outputs,  than the Hirsch index could indicate. Therefore, to scorn upon co-authors in publications, in particular in proceedings, is incorrect. On the contrary, the findings suggest an immediate test of coherence of scientific authorship in scientific policy processes.  
  \end{abstract}
 
  \section{Introduction  }\label{sec:intro}

In recent years, studies of complex systems have become widespread among the scientific community, specially
in the statistical physics one. Many examples, e.g.,  \cite{RMPFortunato,PREMendes,PRESilva}, pertain to social phenomena in general, indicating that physicists have gone far from their traditional domain of investigations   \cite{Auyang,AlbertBarabasi,Boccara,Sornette}. Moreover,  one very modern topics of investigation is the role of measuring (as accurately and objectively as possibly done as in physics)  the value of some scientific production  \cite{PetersenStanleySucci,IvanovDimitrovaIvanovbook}.
  
In \cite{Sofia3a}, it was shown that   a  Zipf-like law
 \begin{equation}   \label{eqZipf}  J \propto 1/r ,
  \end{equation} 
  exists,  between 
the number ($J$) of  joint publications (NJP) of a scientist, called for short "leading investigator" (LI) with her/his coauthor(s)  (CAs); $r $ =1,...   is an integer allowing some hierarchical ranking of  the CAs; $r=1$ being the most prolific coauthor with the PI.  The  number of different coauthors  (NDCA)  is  given by the highest possible rank $r_M$.   Several CAs have often the same NJP with  the LI.  

  It was  observed that  a hyperbolic  (scaling) law   is more appropriate, i.e., 
 \begin{equation}    \label{eq1}
 J =  J_0/r^{\alpha} ,
  \end{equation}
 with $\alpha\neq1$, usually  such that  $\alpha\le1$, and often decreases with the number of CAs or with the number of joint publications,  e.g. when the number of CAs  and when $J$ are   "not  large". $J_0$  is a fit parameter, i.e. there is no meaning to $r=0$.
 
 As the $h$-index \cite{hindex,hirsch10,rousseau06}   "defines"  the {\it core of papers of an author} from the relationship between the number of citations $n_c$ and the corresponding rank $r$ of a paper, through a trivial threshold,  i.e. if $n_c$ $\ge r_c$, then $r_c\equiv h$,  thus
  one is allowed  also to define  the {\it core of coauthors of a scientist} through  a  threshold  \cite{Sofia3a},
 called  the  $m_a$-index,
   \begin{equation} \label{eq2}
 m_a \; \equiv \; r,   \;  \;  \;  \;  as  \;long\;  as\;\; \; \; r\; \le \; J.
  \end{equation}
 This is a specific measure of the core of the most relevant  CAs in a research team,   centered on  the  LI.  In brief, in the $h-$index method,  one implicitly assumes that the number of "important papers" of an author, those which are the most often quoted, allows to measure the impact of a researcher \cite{AlonsoetJOI3.09,Laudel,1-s2.0-S1751157711000630-main,1-s2.0-S1751157711000800-main}. No need to discuss lengthily the $h$-index power, variants, or defects. However, such a citation effect is often due to the activity of  a research team,   centered on  the  LI  \cite{HKretschmer85structuresize,MelinPersson96,Kwok05,[22]}.    In fact, the size and structure of a temporary or long lasting group  is surely relevant to the productivity of an author \cite{1-s2.0-S1751157711000630-main}. 
  In contrast, the $m_a$ index as introduced  measures  the role of coauthors, {\it rather than citations},  to  indicate  the most important coworkers of a LI, allowing to measure the LI team core.
    Technically, one could thus measure the relevant strength of a research group centered on some leader and measure  some impact of research collaboration, e.g.,  on scientific productivity 
\cite{LeeBozeman}.  
 The invisible college  \cite{HK94,Zuccala06invisblcoll}   of a PI would become visible, easily quantified, whence pointing out to some selection in the community.  
     
  Several  other measure definitions  can be deduced, as in the $h$-method, i.e. taking  into account  the  whole surface of the histogram,  i.e.  the cumulated number of joint publications (NJP) 
 \begin{equation}\label{sum}
\Sigma  \equiv \sum_{r=1}^{r_M} J_{r} \;\;  ,
\end{equation}
for the CA with rank $r$ has published $J_r$ publications with the LI.
 A often discussed part of  the histogram is that up to the treshold; it corresponds to the cumulated NJP limited to the core, i.e.
 \begin{equation}\label{suma}
A_{a}   \equiv \sum_{r=1}^{m_a} J_{r}  .
\end{equation}
The notation is reminiscent of  the $A-$index \cite{Jin06,Jin07,JinRE07},  in the Hirsch scientific output measurement method of an author.  Of course, $A_a/\sum$ gives the relative weight of the core CAs in the cumulated NJP. 

Moreover, one can define an   $a_a$-index  \cite{Sofia3a} which measures the  surface below the empirical data of the number of joint publications {\it  till the CA of rank } $m_a$,  normalized to $m_a$, i.e.  
\begin{equation}\label{a_a}
 a_a =\frac{ 1}{m_a} \sum_{r=1}^{m_a} J_{r}\; \equiv \; \frac{A_a}{m_a},
  \end{equation}
  and  similarly  the   index  
  \begin{equation}\label{a_aa}
a_M  =\frac{ 1}{m_a} \sum_{r=1}^{r_M} J_{r}\; \equiv \; \frac{\Sigma}{m_a}.
  \end{equation}
 measured from the $whole$ histogram surface. Obviously,  $A_a/\sum$ $\equiv a_a/a_M$. The notations are similar to those of the $h$-index scheme, where they somewhat measure the average number of citations of papers {\it in the Hirsch core}  \cite{rousseau06}.

Note that the true mean $\mu$ of the $J$ $vs.$ $r$ distributions, i.e. the average NJP  per CA,  is  obtained from 
 \begin{equation}\label{mu}
 \mu =\frac{\Sigma}{  (NDCA)} \; \equiv \; \frac{\Sigma}{r_M}.
\end{equation} 

  In practical terms,   these indirect measures are  attempts to improve the sensitivity of the  threshold forced index in order  to take into account the number of  co-authors  whatever the number of joint publications among  the most frequent coauthors, and introduce a contrast between the most frequent CAs and the less frequent ones.   Indeed,  JP have often a mix of different CAs\footnote{The order of authors is at this level not discussed.}  \cite{1HBPhysA}. It has also been observed in previous  fits, through Eq.(\ref{eq1}), that  unusual  (long or short)   lists of coauthors, as well as the hapax-like CA, i.e. accidental or rare CAs, but with necessarily  large $r$ values,    have much  influence on $J_0$ and $\alpha$, and the resulting $R^2$.   

Moreover, it  is somewhat  commonly accepted that  proceedings papers, e.g. resulting from conference presentations,  have to be distinguished from peer review journal publications.   Miskiewicz  \cite{1JMproceedingscore}  has discussed whether  such different types of publications   have some impact on the core number and on the ranking of CAs.  
For completeness, note that a complementary question was also examined, i.e. whether a  "binary scientific star"-like system implies  some deviation from     Eq.(\ref{eq1}), - the  "binary scientific star"   (BSS) being defined as  the couple formed by a LI and one of his  most frequent CAs \cite{binarystars}.  
 
In the following sections, an amazingly simple relationship is   reported to be found between  $ m_a^{(jp)}  $ and its related value for publications in peer review journals (j) and in proceedings (p),  i.e.  $  m_a^{(j)} +   0.4\;m_a^{(p)} = m_a^{(jp)}  $.         Moreover, another relationship is found concerning the $A_a$ index, i.e. the surface of the $J$ $vs.$ $r$ histogram up to $m_a$, i.e., $A_a^{(j)} + 1.36\; A_a^{(p)}  = A_a^{(jp)} $.  A discussion of other empirical  (linear) relations is presented. The illustrative data  of the  coauthorship features  is quickly  recalled   for  the few published cases, in Sect. \ref{sec:dataanal}.  In Sect. \ref{sec:dataset},  some hint is presented on some origin of the,  surprising (or unexpected), relationship, and for the coefficient values.  The case of anomalous data points is also discussed. Some justification   is based on the empirical power laws $J^{(t)}\propto r^{-\alpha^{(t)}}$,  \cite{Sofia3a},     emphasizing that $\alpha^{(t)}$  depends on the type ($t$)  of publication, even  if only  slightly.  A bonus (?) is found to be the simple power law relationship between the $m_a$ core value and the number of different coauthors, see an Appendix.

 Note that there is at first no reason to  predict  that a simple relationship  will be found between the various quantities here above introduced.  In fact an examination of the distributions led to ambiguous results  \cite{binarystars}.  There is apparently no other previous investigation of this matter.  {\it In fine}, modeling likely requests much more thinking.  Sect. \ref{sec:conclusions} serves as a conclusion   on the respective relevance of different types of publications in "evaluating" a LI and his/her CAs, and  with some suggestion for future work.    Nevertheless, the findings could imply practical considerations on  subsequent measurements of publication activities during a career, -as self-citations might do \cite{Iinamadrid}.

     Note  also that several other so called  $laws$ have been predicted or discovered  about relations between number of authors, number of  publications, number of citations,  fundings, dissertation  production, citations, or the number of  journals or scientific  books,  time intervals, etc.   \cite{egg1}.
     
       \begin{figure}
\centering 
\includegraphics [height=14.0cm,width=14.5cm]
{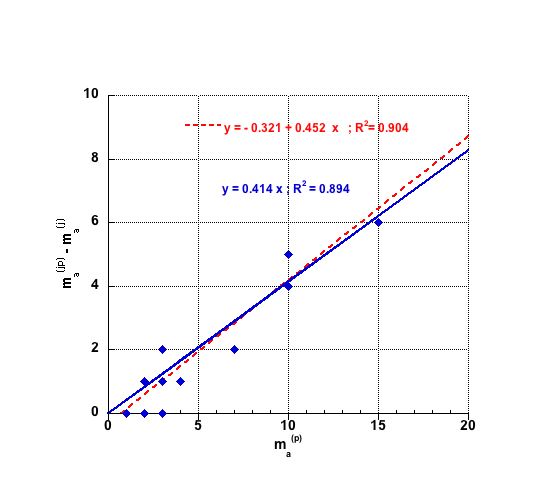}
\caption   { Empirical  proof of the  linear relationship between the total CA core value $m_a^{jp)}$ and the corresponding ones, but distinguishing between peer review journals (j) and "proceedings" (p) papers, i.e. $  m_a^{(j)} +   0.414\;m_a^{(p)} \simeq  m_a^{(jp)}  $;  $R^2=0.894$; the dash line indicates the best possible two parameter linear fit; several data points overlap each other } 
 \label{fig:Plot2y0414x}\end{figure}
 
   \begin{figure}
\centering
 \includegraphics [height=14.0cm,width=14.5cm]{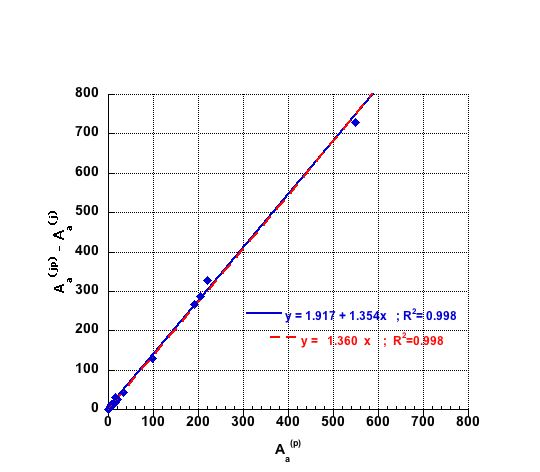}
\caption   { Empirical  proof of  the   linear relationship, $A_a^{(jp)} \simeq A_a^{(j)} + 1.36\; A_a^{(p)}$,  between  $A_a$'s, i.e. the (Number of joint publications-Coauthor rank) histogram $J$ $vs.$ $r$ surface below $m_a$ \cite{Sofia3a},    and the  $A_a$ corresponding values for  peer review journals (j) and "proceedings" (p);  $R^2=0.998$; the dash line indicates the best possible two parameter linear fit; several data points overlap each other  } 
 \label{fig:Plot35Aay136x} \end{figure}

   \begin{figure}
\centering
 \includegraphics [height=14.0cm,width=14.5cm]{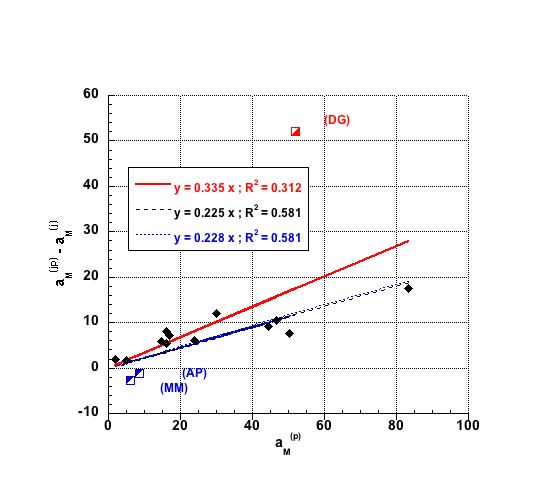}
\caption   {  Empirical  linear relationship, $a_M^{(jp)} \simeq a_M^{(j)} + 0.225\; a_M^{(p)}$, between  the whole $a_M$'s  and the corresponding ones for joint papers with coauthors in  peer review journals (j) and "proceedings" (p);   $R^2\simeq 0.581$; if either the (DG) and the (AP) and (MM) points are not considered in the fit, as being possible "outliers", the relationship  numerical values are slightly modified. N.B. The best possible linear two parameter fits are not shown though have a higher $R^2$, see text for values, but the abscissa at the origin can hardly be interpreted} 
 \label{fig:Plot9aM3cases} \end{figure}

 \begin{figure}
\centering
 \includegraphics [height=14.0cm,width=14.5cm]{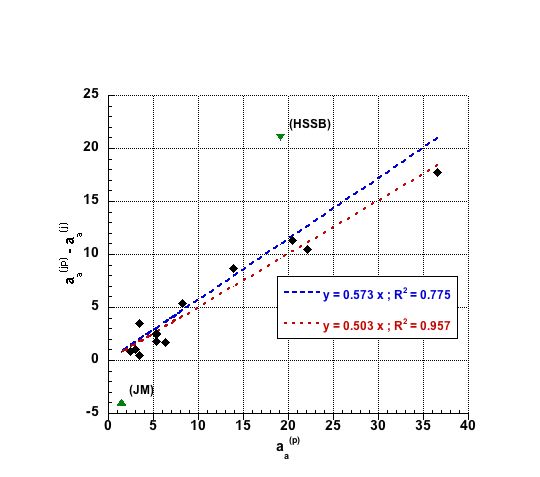}
\caption   {  Empirical  proof of the  linear relationship, $a_a^{(jp)} \simeq a_a^{(j)} + 0.503\; a_a^{(p)}$, between   $a_a$ values  resulting from  the (Number of joint publications (jp)-Coauthor rank) histogram, $J$ $vs.$ $r$,  surface  below the core value $m_a$,    and that of   peer review journals (j) and "proceedings" (p); $R^2$ = 0.957, when  (JM)  and (HSSB) points are not considered, as being  outliers, see text;   inclusion of such points are shown by the blue dash line;  the best linear two parameter fits are not shown because the abscissa at the origin can hardly be interpreted } 
 \label{fig:Plot214aa503573} \end{figure} 


   \section{Data sample }   \label{sec:dataanal}
 
For the following study and discussion  the same LIs  as those investigated in previous publications   \cite{Sofia3a,1HBPhysA,1JMproceedingscore,binarystars} are considered. The LIs span a large range of scientific  research topics, though in statistical physics mainly. They are  mentioned by their initials. Most of them are males, except two. They come from mainly Poland, 7,  i.e. RW, JMK, AP,  KSW, JM, and MM; 4 are from the "western world", HES, DS, MA, and PC. They are  half  several senior (JMK, AP, HES, DS, MA, PC) and  half rather junior scientists. In previous reports   \cite{Sofia3a,1HBPhysA,1JMproceedingscore,binarystars} , their publication list has been summarized and is thus not recalled here. Beside the LIs,   4 BSS cases \cite{binarystars}, i.e. so called  HSSH, HSSB, MARC  and   MANV has been made for further  completing the up to-day rather rare data. This leads to examine  15 cases. {\it A priori}, the data does not seem to be specifically biased. 

The best power law fits, through Eq.(2), $\alpha$ and $m_a$ value, and the distribution main statistical characteristics, i.e., the mean  $\mu$ and $r_M$, are  given in Table 1. This  well illustrates the similarity in behavior, but points to differences to be examined next in more detail.  One may expect, from a general  point of view, that the subsets, i.e. joint publications in peer review journals (j) and in proceedings (p),  might have some influence on characteristics of those for the whole (jp) set  since they form the structure. However, the problem is highly non linear in essence, since the "rank" is not a usual variable. Nevertheless, in line with modern statistical analysis, and in order to detect some substructure, several fits can be attempted, i.e. power law, exponential, logistic, ... and polynomial, the most simple being the linear one. 

The deduced values of  $\sum$, $A_a$, $a_M$ and $a_a$ are given for the three sets, i.e. (jp), (j) and (p), in Table 2.  The empirically found laws are presented in Figs. 1-4, for  the $m_a$, $A_a$,  $a_M$,  $a_a$ quantities. 
 
A simple relation is found between the  core measures,  i.e.
 
     \begin{equation} \label{m_aeq}
 m_a^{(j)} \;+  \; 0.414\;m_a^{(p)} = \;m_a^{(jp)} 
   \end{equation}
with a high regression coefficient $R^2$, i.e. $\sim0.894$, and between the histogram surfaces below the corresponding core measures
  \begin{equation} \label{A_aeq}
   A_a^{(j)}\; + \;1.36\; A_a^{(p)}  = \;A_a^{(jp)}  
   \end{equation}
with a very high regression coefficient $R^2$, i.e. $\sim0.998$, as seen in  Figs. \ref{fig:Plot2y0414x}-\ref{fig:Plot35Aay136x} respectively.   In both cases, a classical two parameter linear fit has been attempted. It has been found that   $m_a^{(jp)} = m_a^{(j)} +   0.452\;m_a^{(p)} -0.321 $ and $ A_a^{(jp)}= A_a^{(j)} + 1.354\; A_a^{(p)} + 1.917   $ respectively, with the corresponding $R^2$ values being equal to 0.904 and 0.998.  Such fits are shown in Figs. \ref{fig:Plot2y0414x}-\ref{fig:Plot35Aay136x}. 
 
 Let it be observed that one has necessarily 
    \begin{equation} \label{sumeq}
      \Sigma ^{(j)} \;+  \; \Sigma ^{(p)}  \; = \; \Sigma ^{(jp)} 
   \end{equation}
which is nothing else that  a normalization condition  on the NJP,  as exemplified  in Table 2.

Subsequently, $a_M$ has been examined. The result is reported in Fig. \ref{fig:Plot9aM3cases}. One finds
     \begin{equation} \label{a_Meq}
    a_M^{(j)} \;+  \;0.225 \; a_M  ^{(p)}  \; = \; a_M  ^{(jp)}\;\;, 
   \end{equation}
 but with a low $R^2$ $\simeq 0.581$. More discussion, stemming from  the presence of anomalous data,  so called "outliers", is found in Sect. \ref{sec:dataset}.  
 
  Finally,  a fine linear fit is found for $a_a$ as seen in Fig. \ref{fig:Plot214aa503573} 
     \begin{equation} \label{a_aeq}
    a_a^{(j)} \;+  \;0. 503 \; a_a \;  ^{(p)}  \; = \; a_a \;^{(jp)}\;\;,
   \end{equation}
 with $R^2 \simeq 0.957$,   even with the presence of anomalous  (outlier) data,   
 as discussed   in Sect. \ref{sec:dataset}.

  \section{  Discussion  }\label{sec:dataset}
  
  \subsection{Outliers}

   First let us discuss the cases of $a_M$ and  $a_a$, before introducing an 
    estimate of the numerical coefficients. Visual inspection shows that there are a few data points looking like outliers\footnote{Interesting comments on outliers can be found in \cite{WCS1.09.57,WCS5.13.30}}
     in 
   Fig. \ref{fig:Plot9aM3cases} and
    Fig. \ref{fig:Plot214aa503573}. They are called (DG), (AP), and (MM) on one hand 
   and (JM) and (HSSB) on the other hand.

  For the $a_M$ case,    Fig. \ref{fig:Plot9aM3cases},  it is remarked that (DG)  has a specially high number of  CAs, i.e. 93,  having only one ("proceedings") joint publication with DG. This stems from some work by DG  in high-energy physics before he  turned towards statistical mechanics research. If  such a data point is included in a proportionality fit between $a_M^{(jp)}-a_M^{(j)}$ and $a_M^{(p)}$, one obtains a proportionality coefficient equal to 0.335, and a $R^2$ value  equal to 0.312. Concerning (AP)  and (MM), it is observed that  $a_M^{(jp)}-a_M^{(j)}\le 0$, which may at first  appear awkward and unattractive. Note that the respective values of $\sum$ and $m_a$ seem "reasonable".  However, the origin of the negative value is likely attributable to the fact that AP and MM have very few "proceedings" papers, having  mainly concentrated their research  output into peer review journals, and few papers resulting from scientific meetings. In some sense, through these two authors, one point out to the effect of duplicate-like papers.
   
   The comment on the"(DG) effect" implies that one should deduce that the proportionality coefficient is likely to be dependent on  the research field.  On the other hand,  the  comment on the "(AP)-(MM) effect "indicates that one should allow for a negative value of  $a_M^{(jp)}-a_M^{(j)}$, and propose that the coefficient to be accepted is 0.228 rather than 0.225.
   
   For completeness, a classical two parameter linear fit  has been made either taking into account all data points, or removing outliers. The following  relations have been obtained,  for  the $a_M$
   \begin{itemize}
   \item  taking into account all data points,
   $y= 0.255 + 0.33x$, with $R^2=0.349$,
   \item without the (DG), (AP) and (MM) points, 
   $y= 2.964 + 0.164x$, with $R^2=0.769$,
   \item without the (DG) point, 
   $y=1.232+0.197x$, with $R^2=0.718$.
   \end{itemize}

For the $a_a$ cases,  see Fig. \ref{fig:Plot214aa503573},   HSSB  is seen to have a very high   $a_a^{(jp)}-a_a^{(j)}$  positive value, while for JM , one obtains  $a_M^{(jp)}-a_M^{(j)}\le0$.  It is fair to emphasize that there is a similarity between $a_a$  and  $a_M$ cases. However,  the  deductions originate from  different surfaces, i.e. below the cores in the $a_a$ cases,  to be  contrasted with the whole surfaces for the $a_M$ cases. 

Since HSSB have almost equal NJP,  with their main CAs,  in either $p$ or $j$ set, one might wonder if these are duplicate results, since they imply the same and main CAs. The case of JM shows that this is not a duplicate finding: indeed, on the contrary, JM  has  very few $p$-type publications with his/her main CAs. Moreover, the different  "behavior" of HSSB and JM enlightens the fact that JM has rather a concentration of scientific output in peer review journals with his/her  main CAs  (like  for AP and MM, in fact). 

   For completeness, a classical two parameter linear fit  has been made taking into account all data points, or removing outliers. The following  relations have been obtained,
    for  the $a_M$   \begin{itemize}

   \item  taking into account all data points,
   $y= -0.671+ 0.608x$, with $R^2=0.775$,
   \item without the (HSSB)  and (JM) points, 
   $y=-0.07+0.507x$, with $R^2=0.957$.
   \end{itemize}

Note that one should not be impressed by the respective $R^2$ values, - obviously depending on the number of data points, and the fact that these are two parameter fits, - in contrast to the  values given here above and in either Eq.(\ref{a_Meq}) or Eq.(\ref{a_aeq}).

   The positive or negative value of the abscissa at the origin can be attributed to the fact that the resulting combination (jp) between the (j) and (p) set is {\it a priori} highly non-linear. Indeed the various ranks do not sum up, since a CA in one set may appear at two rank values totally unrelated to the resulting rank for the (jp) set. 

Finally, in all cases, one may deduce that these (outlier-like) results arise from different scientific (or other) behavior of the respective scientists, but this discussion is outside the realm of the present paper.

 \subsection{Theoretical estimates}
 
 The numerical proportionality coefficients, as well as those resulting from a two linear parameter fit,   can be discussed, going to the continuum limit for the respective histograms. Indeed, within a continuum approximation, one has
 
 \begin{equation}\label{continuum}
 \sum_{r=1}^{r}  J_r  \rightarrow   \int_{1}^{r} J(r)  \;dr \equiv  \int_{1}^{r}\frac{1}{r^{\alpha}}\,dr  = J_0\;[r ^{(1-\alpha)}-1].
 \end{equation}
  
 Consequently,
  \begin{equation}\label{continuumSUM}
 \sum_{r=1}^{r_M}  J_r \;\equiv\; \Sigma \rightarrow \;  =\;J_0\;[r_M ^{(1-\alpha)}-1],
 \end{equation}
  \begin{equation}\label{continuumAa}
A_a \equiv  \sum_{r=1}^{m_a}  J_r  \; \rightarrow \;   =  J_0\;[m_a^{(1-\alpha)}-1],
 \end{equation}
  \begin{equation}\label{continuumaM}
\frac{ \sum}{m_a}  \; \rightarrow   \;    = \frac{J_0}{m_a}\; [r_M ^{(1-\alpha)}-1],
 \end{equation}
  \begin{equation}\label{continuumaa}
\frac{A_a}{m_a} \;  \rightarrow  \;     = J_0\;[m_a ^{-\alpha}-m_a^{-1}].
 \end{equation}
 
    Recall that these quantities depend both on the type of publication set and on the LI.    For example, for one specific LI, and some publication set, one has from Eq.(\ref{continuumAa}),
      \begin{equation}\label{continuumaaLI}
      (A_a/J_0) +1 = m_a^{1-\alpha},
     \end{equation}
 while from Eq.(\ref{continuumSUM}), one finds
        \begin{equation}\label{continuumSUMLIeq}
     [\;\sum/J_0)\;] +1 = r_M^{1-\alpha}.
     \end{equation}

\begin{figure}
\centering
 \includegraphics [height=14.0cm,width=14.5cm]{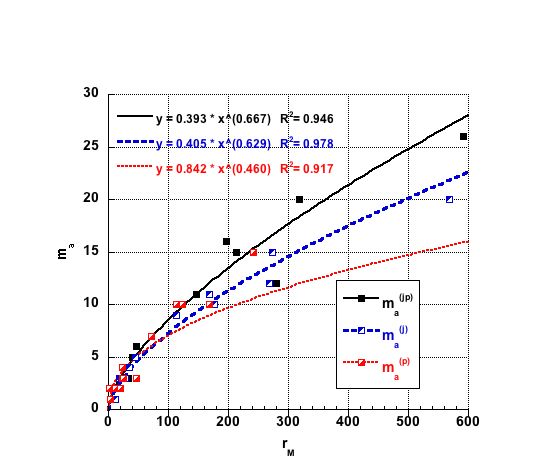}

\caption   {  Empirical  proof of the  power law relationship, $m_a \simeq r_M^{\beta} $,   resulting from  the (Number of joint publications (jp)-Coauthor rank) histogram, $J$ $vs.$ $r$,  surface   and that for   peer review journals (j) and "proceedings" (p), when the  (DG) point is  not considered, as being   an outlier  } 
 \label{fig:Plot4allamvsrMnoDG} 
\end{figure}

\begin{figure}
\centering
 \includegraphics [height=14.0cm,width=14.5cm]{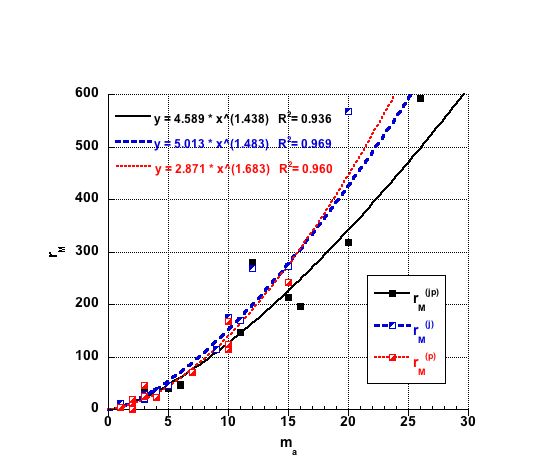}

\caption   { Empirical  proof of the  power law relationship, $r_M \simeq m_a^{\gamma}$,  when  the outlier (DG) point is  not considered, for the cases of  peer review journals (j) and "proceedings" (p), and the total (jp) number of joint publications } 
 \label{fig:Plot16allrMvsmanoDG} 
\end{figure}

Observe that 
  \begin{equation}\label{frac}
     \frac{\sum + J_0}{A_a+ J_0} = (\frac{r_M}{m_a})^{(1-\alpha)}.
     \end{equation}
     for each LI and each type of scientific publication.   Some elementary, but very tedious algebra, can follow. From Eq.(\ref{frac}), one can extract $\sum$, and rewrite explicitly Eq.(\ref{sumeq}), in order to obtain a linear relationship between the three $A_a$  quantities, such that one  can write 
     
  \begin{equation}\label{linAa}
  A_a^{(jp)} = \Lambda_j \;   A_a^{(j)} +   \Lambda_p\;   A_a^{(p)}  +  [\;\;....\;\;],
    \end{equation}
where  each $\Lambda$ and [$\;\;...\;\;$]  are functions of $r_M$ and $m_a$.  Moreover, as shown in Appendix,  one has a (surprisingly simple) power law relationship between $m_a$ and $r_M$, i.e. $m_a\;=\;v\;r_M^{\beta}$, 
 (or $r_M\;=\;u\; m_A^{\gamma}$),
see Figs. \ref{fig:Plot4allamvsrMnoDG}-\ref{fig:Plot16allrMvsmanoDG}. Therefore, one can evaluate the ratio $\Lambda_p / \Lambda_j$  for  the various cases; it is about $v_p/v_j \sim (2.9/5.0)^{-(2/3)}$ $\;$ $\sim 1.44$, not too far from 1.36. The complicated  [$\;\;...\;\;$]  term can be roughly estimated, according to the various numerical values. It is found rather small with respect to the other terms, whence corroborating the finding in Eq.(\ref{A_aeq}). 

To "verify" Eq.(\ref{m_aeq}) is more subtle and complicated. Indeed, one has to start from Eq.(\ref{sumeq}), in which one substitutes each  Eq.(\ref{continuumSUM}). This leads to a highly non linear relationship of the type

  \begin{equation}\label{linSum}
 M_a^{(jp)} = \Omega_j \;   M_a^{(j)} +   \Omega_p\;   M_a^{(p)}  +  [\;\;....\;\;],
    \end{equation}
    where
    \begin{equation}
    M_a\; \equiv \; m_a^{\gamma(1-\alpha)}.
 \end{equation}
 To extract  a linear relationship, analytically, like $ m_a^{(jp)} = \omega_j \;   m_a^{(j)} +   \omega_p\;   m_a^{(p)}  +  [\;\;....\;\;],$ similar to  Eq. (\ref{m_aeq})  seems  feasible only if $ \gamma(1-\alpha)\equiv (1/k)$,  where $k$ is   an integer.  For the sake of argument, taking $\gamma =3/2$ and $\alpha=1/3$, thus $k=1$, one can estimate according to the values in Tables 1-2   and those in the Appendix that  the ratio $ \omega_p\; /\omega_j\; \sim J_0^{(p)} /J_0^{(j)} \; \simeq0.5$,  not too far from 0.4 in  Eq. (\ref{m_aeq}). 
 
   \section{Conclusions} \label{sec:conclusions}

 In summary,  recall that the core $m_a$ of coauthors (CAs)  of a leading investigator (LI) is well defined through a criterion similar to the $h$-index.  Through a $J$ $vs.$ $r$ histogram,   one describes a  CA "impact"  according to the number of his/her joint publications $J$ with a  LI.  It is usually considered that  scientific publications in proceedings differ, in various ways, "values",  from those in peer review journals. This belief was here above questioned. In fact, one can distinguish the core of coauthors of a LI, according to the type of  joint publications.  Next,  it was  wondered whether the  relative hierarchy in estimating the value of publications in  journals or proceedings can be carried to the core of co-authors. Finally,  the question was raised: "is there any numerical proof of the usual belief in a qualitative difference"?
 
 Visually, it appears that the hierarchy exists, i.e. $m_a^{jp} > m_a^{j}  > m_a^{p}$, but, somewhat surprisingly,  the relationship turns out to have a simple analytic form:   $  m_a^{(j)} +   0.4\;m_a^{(p)} = m_a^{(jp)}  $.  Moreover, another relationship is found concerning the $A_a$ index, i.e. the surface of the $J$ $vs.$ $r$ histogram up to $m_a$, i.e., $A_a^{(j)} + 1.36\; A_a^{(p)}  = A_a^{(jp)} $. These linear relationships also  hold  for  subsequently derived  measures. 
 
 The findings have been illustrated gathering data for a dozen or so LIs, and for 4 couples, i.e. publications in which  a LI is systematically with a specific CA.  Even though the data, about scientists  working in the research field of statistical physics, is of finite size,  the result does not seem to be biased.   A $\chi^2$ test  indicates the reliability of the  found features. Yet, in  one case, a LI, having previously worked in high energy physics and having many (93) CAs for  one  publication, of the (p)-type, some anomalous behavior occurs.  This outlying feature should not  distract from the main  findings. 
 
 The main text analysis  points out to the interest of the measures of CA cores according to types of publications.  The results  of course suggest to investigate other scientific domains. In fact, it can be done through a bonus, the discovery that $m_a$ is tied  with  the maximum number of CAs  of a LI,  i.e. $r_M$, through a simple power law. In so doing, one may observe that outliers are easily found   \cite{WCS1.09.57,WCS5.13.30}, - thus removing them leads to a large increase in the $\chi^2$ coefficient, {\it in fine} giving much weight to the interest of such numerical findings. It is also obvious that the analysis is rather simple for policy makers, - when the list of publications is available.
 
 Theoretical work, whence explanations, have been shown not to be trivial, because the systems are highly non-linear. A CA might appear in one type ($j$ or $p$) of publications, but not in  the other type. However, as it is more frequently seen, a CA appears in both types of publication.However, it might be at a quite different ranking. This is the case when a CA does not have "many" publications with a LI.  In fact, the ranking of some CA in some type of publication does not simply add up with the ranking in another type of publications.  A CA ranking  is usually not conserved, but depends on the publication type. Whence, there is {\it a priori } no reason why a linear relationship  should be found between such measures. 
 
 However, there is one "normalizing condition", in general not appreciated, which imposes further consideration of such co-authorship measures: the surface of the histogram $J$ $vs.$ $r$  for the whole set of publications is necessarily the sum of the surfaces for the two types of publications. Thus, since a the continuum limit,   $J \propto  1/r^{\alpha}$, one can measure such surfaces as a function of $\alpha$ and  $r_M$, and later derive an estimate of the numerical coefficients  obtained through empirical fits.
 
 Finally, one should not simply consider that these are numerical games.  They lead to remarkable proportionality measures of CA roles. It appears that the number of CAs "of interest" for measuring the core of CA of a LI is mainly arising from the  joint publications in peer review journals. Indeed, only about  (0.4/1.4 $\simeq$) 30\% stem from "proceedings". Similarly,  it appears that the "contribution" to the number of joint publications by the main CAs is about  50\% of the whole.  
 
 This   implies to elaborate practical considerations on  subsequent measurements of publication activities during a career, on the role and effects of coauthors  \cite{8,Perssonetal04,galam}.   The case of outliers is also of interest, since it carries some weight in estimating $\alpha$ and subsequent numerical coefficients.  
 
 These  simple relations imply that  coauthors  are a "more positive measure"  of a principal investigator role, in both types of scientific outputs,  than the Hirsch index  which barely counts the number of citations independently of the co-author (number nor  {\it a fortiori} rank). Therefore, to scorn upon co-authors in publications, in particular in proceedings, is  highly incorrect. On the contrary, the findings suggest an immediate test of coherence of scientific authorship in scientific policy processes.  This could imply many practical considerations on the role of CAs with respect to a LI and on the respective roles of different types of publications in "measuring" a LI team work. A discussion of criteria based on the above for estimating, e.g the financing of a LI or a team,  is outside the realm of the present paper. Nevertheless, with softwares actually  available to policy makers, the development and application of such findings should be easily possible.

 Note three final points: (i) each rank-frequency form, like Eq.(1) or Eq. (2), has an equivalent size-frequency one \cite{Egghebook05}. One should become curious about whether similar equalities hold for the size-frequency cases;
(ii)  the above considerations suggest to investigate if similar relationships exist for the $h$-index, distinguishing between $h^{j}$, $h^{p}$, and $h^{jp}$,  and to draw {\i ad hoc} conclusions; (iii)
 complex systems do not necessarily lead to find non linear laws.
 
   \vskip0.5cm
 {\bf Acknowledgements}
 This paper is part of scientific activities in COST Action  TD1306
"New Frontiers of Peer Review (PEERE)"
 
  \bigskip
  
 {\bf Appendix} \vskip0.3cm {\bf Other empirical laws: $m_a$ $vs.$ $r_M$ and $r_M$ $vs.$  $m_a$, with or without outliers}  \label{App:marMrMma}
 
  \bigskip
For the theoretical estimation of the numerical values in Eqs.(\ref{m_aeq})-(\ref{a_aeq}),  in particular in order to  proceed beyond Eq.(\ref{frac}), it appears that it is useful to find  whether a relationship exists between $r_M$ and $m_a$ on average.

The most simple power law fits for  $m_a$  $vs.$ $r_M$, i.e. $m_a\simeq r_M^{\mu}$,   give,   taking into account all data points,

$m_a^{(jp)}$ = 0.392 $\;\;$  [$r_M^{(jp)}]^{(0.645)}$  , with $R^2$= 0.893 

$m_a^{(j)}$  $\;$ =  0.405 $\;\;$ [$r_M^{(j)}]^{(0.629)}$   , with  $R^2$= 0.978

$m_a^{(p)}$  $\;$ =  0.904 $\;\;$ [$r_M^{(p)}]^{(0.415)}$   , with  $R^2$= 0.760

but lead to

$m_a^{(jp)}$ = 0.393 $\;\;$ [$r_M^{(jp)}]^{(0.667)}$   , with  $R^2$= 0.946   

$m_a^{(j)}$  $\;$ =  0.405 $\;\;$ [$r_M^{(j)}]^{(0.629)}$  , with $R^2$= 0.978  

$m_a^{(p)}$  $\;$ =  0.842 $\;\;$ [$r_M^{(p)}]^{(0.460)}$   , with  $R^2$= 0.917

 when the outlier (DG) data point is not taken into account, i.e. roughly speaking $m_a \simeq 0.5 \;  m_a^{\gamma}$, with $\gamma \simeq $ 1/2 or 2/3. Similarly, the best power law fits for  $r_M$ $vs.$ $m_a$ give, when taking into account all data points,

$ r_M^{(jp)}$ = $\;\;$8.691$\; $ $[ m_a^{(jp)} ]^{(1.175)}$, with  $R^2$= 0.893   

$r_M^{(j)}$$\;\;$ =  $\;\ $5.013$\;\;$ $[m_a^{(j)}]^{(1.483)}$, with  $R^2$= 0.969   

$r_M^{(p)}$ $\;\;$=  $\;\;$4.261$\;\;$ $[m_a^{(p)}]^{(1.507)}$, with  $R^2$= 0.867   

but  become

$  r_M^{(jp)}$ = $\;\;$ 4.589 $\;\;$ $[m_a^{(jp)}]^{(1.438)}$ , with  $R^2$= 0.936   

$r_M^{(j)}$ = $\;\;$  5.013 $\;\;$ $[m_a^{(j)}]^{(1.483)}$ , with $R^2$= 0.969    

$r_M^{(p)}$= $\;\;$ 2.871 $\;\;$ $[m_a^{(p)}]^{(1.683)}$ , with   $R^2$= 0.960   
 
 when the outlier (DG) is not taken into account, i.e. roughly speaking $r_M \simeq 4.5 \;  m_a^{\beta}$, with $\beta \simeq $ 3/2 or 5/3. The "no (DG)"  cases are shown in Figs. \ref{fig:Plot4allamvsrMnoDG}-\ref{fig:Plot16allrMvsmanoDG}  for illustration of the findings. Note the large $R^2$ values.
 
 \bigskip

 {\bf Acknowledgements}  \bigskip
 
 Thanks to  H. Bougrine 	and  J. Miskiewicz   for discussions on their      \cite{1HBPhysA} and \cite{1JMproceedingscore}   respectively.  Special thanks to  J. Miskiewicz  for providing unpublished data on polish scientists, with career comments.

 \begin{table}\label{table1majpmajmap}\begin{center}
      \begin{tabular}{|c|c|c|c|c|c|c|c|c|c|c|c| c|c|c|    c|c|c| 
      }
  \hline
   
&
&$m_a^{(jp)}$&$m_a^{(j)}$&$m_a^{(p)}$
&$\alpha^{(jp)}$&$\alpha^{(j)}$&$\alpha^{(p)}$
&$\mu^{(jp)}$&$\mu^{(j)}$&$\mu^{(p)}$
&$r_M^{(jp)}$& $r_M^{(j)}$&$r_M^{(p)}$
\\
\hline
 \hline    HES &&26&20&15 &1.135&0.999&1.045&6.569&4.67&5.136&592&568&242    \\
 \hline
  \hline     DS  &&  12&12&3&0.796&0.535&0.688&2.725&2.578&1.565&280&268&46   \\
 \hline
 \hline  MA &&20&15&10&1.102&1.029&0.86&4.872&3.865&3.041&319&273&168   \\ 
\hline 
 \hline  PC &&4&3 &3&0.87&0.94&0.67&3.097&2.684&2.0&302&129&24   \\ 
\hline 
 \hline     RW&& 6 &4 &4 &0.743 &0.767&0.561&2.75 &1.94 &2.71 &46&34&23    \\
 \hline
 \hline     JMK&&5&4&3&0.787&0.702&0.618&2.707 &1.714 &2.04&41&35&25    \\
 \hline
 \hline     AP &&6 &5&2&0.94&0.64&0.89&2.872&2.622&1.5455&47&45&11   \\
 \hline
  \hline    DG&&2&2&2 &0.547&0.755&0.239&1.13&1.75&1.05&104&7&99    \\
 \hline
  \hline    KSW&&3&3&1 &0.715&1.255&0.594&2.13&2.0&1.67&21&21&3    \\
 \hline
 \hline    JM && 2  & 2  & 1 &0.63&0.67& 0.67&1.75 &1.71&1.33&14&12&1     \\ 
 \hline
 \hline   MM&& 3  & 2 & 2 &0.536&0.428&0.521&1.515&1.3125&1.45&33&16&20   \\
 \hline
 \hline   HSSH && 16  & 11  &10 & 1.074&0.934&0.974&5.602 & 3.87 &3.895&196&169&114   \\ 
 \hline
 \hline   HSSB && 15 & 10  &10 &1.064 &0.922&0.969 &5.104&3.549&3.766&214&176&114     \\ 
 \hline
 \hline    MARC&& 11  & 9  &7 &0.985&0.893&0.887&3.81&3.07&2.958 &147&114&71     \\ 
 \hline
  \hline    MANV&& 5 & 3  & 3&0.835&0.755&0.67&2.60&2.143&1.833&40&28&24     \\ 
 \hline
\end{tabular}  \end{center}
\caption{Summary of direct data values for 11 LIs and 4 BSSs: $m_a$  is the core measure \cite{Sofia3a};  
$\alpha$ is the exponent of the empirical power law, Eq. (\ref{eq1}); $\mu$ is the mean of the distribution ($J$ $vs.$ $r$);  $r_M$ the total number of different  CAs (NDCA); always distinguishing among of joint publications,  the  total (jp) sum,   the   journals (j) and the  "proceedings" (p) }
\end{table}

 \begin{table}\label{table2majpmajmap}
      \begin{tabular}{|c|c|c|c|c|c|c|c|c|c|c|c| c|c|c|    c|c|c| 
      }
  \hline
   
&
&$a_M^{(jp)}$&$a_M^{(j)}$&$a_M^{(p)}$
&$\Sigma^{(jp)}$&$\Sigma^{(j)}$&$\Sigma^{(p)}$
&$A_a^{(jp)}$&$A_a^{(j)}$&$A_a^{(p)}$
&$a_a^{(jp)}$& $a_a^{(j)}$&$a_a^{(p)}$
\\
\hline
 \hline    HES &&149.6&131.95&83.3&3889&2639&1250 &1625&895&549&62.5&44.75&36.6    \\
 \hline
  \hline     DS  && 63.58&57.58&24&763&691&72&259&229&16&21.58&19.08&5.33   \\
 \hline
 \hline  MA &&77.85&70.33&50.2&1557&1055&502&810&482&221&40.5&32.13&22.1   \\ 
\hline 
 \hline  PC &&24.75&17&16&99&51&48&46&29&16&11.5&9.67&5.33  \\ 
\hline 
 \hline     RW&&21.5&16 &16.25 & 129 & 64 & 65 &64 &21 &33 &10.67&10.2&8.25  \\
 \hline
 \hline     JMK&&22.2&15&17&111&60&51&39&21 &16 &7.8&5.25&5.33    \\
 \hline
 \hline     AP &&22.5 &23.6&8.5&135&118&17&64&51&7&10.67 &10.2 &3.5  \\
 \hline
  \hline    DG&&59&7&52 & 118&  14& 104&14&7&7&7&3.5&3.5   \\
 \hline
  \hline    KSW&&16.33&14.67&5 &49 &44&5&18&15&3&6&5&3   \\
 \hline
 \hline    JM &&13.5 & 11.5  & 2 & 27&23& 2 & 12 &10&3&6&5&3  \\ 
 \hline
 \hline   MM&& 16.67 & 10.5& 14.5& 50&21&29&16&7&8&5.33&3.5&4   \\
 \hline
 \hline   HSSH &&68.625&59.45&44.4& 1098&654&444&524&236&204&32.75&21.45&20.4     \\ 
 \hline
 \hline   HSSB &&72.53& 62.1  &46.7&1088&621&467&469&202&191&31.27&20.2&19.1     \\ 
 \hline
 \hline    MARC&&50.91&38.89  &30 & 560&350&210&280&151&97&25.45&16.778&13.86     \\ 
 \hline
  \hline    MANV&&20.8& 20 &14.667& 104&60& 44  &52&26&19&10.4&8.667&6.33     \\ 
 \hline
\end{tabular} 
\caption{   Summary of indirect data values for 11 LIs and 4 BSSs:  $a_M\equiv\Sigma/m_a$, where $m_a$  is the CA core measure;   $\Sigma$ is the surface below the histogram ($J$ $vs.$ $r$), i.e. TNCA; $A_a$ is the surface below the  $J$ $vs.$ $r$ histogram,  limited to the core value   $m_a$;  $a_A\equiv A_a/m_a$; each  value  for  the  total (jp)  number  of joint publications, journals (j) or  "proceedings" (p) }
\end{table}


\bigskip
     
      \begin{thebibliography}{0}
      
          \bibitem{RMPFortunato}     Castellano, C.,  Fortunato,  S.,  and  Loreto, V.,  Statistical physics of social dynamics, {\it  Rev. Mod. Phys. 81}, 591 (2009).
     
         \bibitem{PREMendes}  Mendes, R.S., Ribeiro, H.V., Freire, F.C.M.,   Tateishi, A.A., and Lenzi, E.K., Universal patterns in sound amplitudes of songs and music genre,  {\it Physical Review E 83}, 017101 (2011).
            
     \bibitem{PRESilva} da Silva, R.,  Vainstein, M. H.,  Gon{\c{c}}alves, S.,  and Paula, F.S.F., Anomalous diffusion in the evolution of soccer championship scores: Real data, mean-field analysis, and an agent-based model, 
 {\it Physical Review E 88}, 022136 (2013).
      
          \bibitem{Auyang}  Auyang, S.Y., Foundations of complex-systems (Cambridge University Press, Cambridge, 1998). .
      
         \bibitem{AlbertBarabasi}   Albert, R. and Barabasi,  A.-L.,  Statistical mechanics of complex networks,  {\it Rev. Mod. Phys. 74}, 47 (2002).
      
           \bibitem{Boccara}   Boccara, N., Modeling complex systems (Springer-Verlag, New York, 2004).
      
         \bibitem{Sornette}   Sornette, D., Critical phenomena in natural sciences (Springer-Verlag, Berlin, 2006).
   
          \bibitem{PetersenStanleySucci} Petersen, A. M.,  Stanley, H. E., and    Succi, S., Statistical regularities in the rank-citation profile of scientists, {\it Scientific reports 1}, (2011).
 
       \bibitem{IvanovDimitrovaIvanovbook} Ivanov, N.K., Dimitrova, Z.I.,  and Ivanov, K.N., Science dynamics and scientific productivity,  (Vanio Nedkov, Sofia, 2014).
   
    \bibitem{Sofia3a}  Ausloos, M. ,  A scientometrics law about co-authors and their ranking. The co-author core.  {\sl Scientometrics 95}, 895-909 (2013). 
    
\bibitem{hindex}     Hirsch, J.E.  , An index to quantify an individual's 
scientific research output.  {\it Proceedings of the National Academy of Sciences USA  102}, 16569-16572   (2005).  

\bibitem{hirsch10}     Hirsch, J.E.  , An index to quantify an individual's scientific research output that takes into account the effect of multiple coauthorship.  {\sl Scientometrics  85},   741-754    (2010).  
 
 \bibitem{rousseau06}  Rousseau, R. ,  New developments related to the Hirsch index.
 {\sl Science Focus 1},  23-25 (2006).

  \bibitem{AlonsoetJOI3.09} Alonso, S., Cabrerizo, F. J., Herrera-Viedma, E., \& Herrera, F.,  h-Index: A review focused in its variants, computation and standardization for different scientific fields. {\it Journal of Informetrics} 3, 273-289  (2009).
  
   \bibitem{Laudel} Laudel, G.,  What do we measure by co-authorships? In M. Davis \& C. S. Wilson (Eds.) Proceedings of the 8th International Conference on Scientometrics and Informetrics (pp. 369-384). Sydney, Australia: Bibliometrics \& Informetrics Research Group  (2001).     

  \bibitem{1-s2.0-S1751157711000630-main}  Abbasi, A.,   Altmann, J.,  \&  Hossain, L. ,   Identifying the effects of co-authorship networks on the performance of scholars :  A correlation and regression analysis of performance measures and social network analysis measures   {\sl  Journal of Informetrics 5}, 594-607  (2011). 
  
     \bibitem{1-s2.0-S1751157711000800-main}  Chien Hsiang Liao, \& Hsiuju Rebecca Yen,Quantifying the degree of research collaboration: A comparative study of collaborative measures,  {\sl  Journal of Informetrics 6},  27-33   (2012). 
            
    \bibitem{HKretschmer85structuresize} Kretschmer, H., Cooperation structure, group size and productivity in research groups. {\sl Scientometrics  7} ,  39-53 (1985). 

   \bibitem{MelinPersson96}  Melin, G.  \&   Persson,O. , Studying research collaboration using co-authorships.  {\sl Scientometrics} 36, 363-377  (1996). 

 \bibitem{Kwok05}    Kwok, L.S., The White Bull effect: abusive coauthorship and publication parasitism.  {\sl  Journal of Medical Ethics  31},  554-556  (2005).  
 
 \bibitem{[22]} McDonald, Kim A.,Too Many Co-Authors?,  {\sl Chronicle of Higher Education,  41} (33),  35-36. 
 
 \bibitem{LeeBozeman}   Lee, S.,  \&   Bozeman, B.     (2005). The impact of research collaboration on scientific productivity,  {\sl Social Studies of Science 35},  673-702   (1995). 

 \bibitem{HK94}   Kretschmer, H.,  Coauthorship networks of invisible colleges and institutional communities.  {\sl Scientometrics} 30, 363-369   (1994).
 
 \bibitem{Zuccala06invisblcoll}  Zuccala, A., Modeling the invisible college.   {\sl  Journal of the American Society for Information Science and Technology 57} (2),  152-168     (2006).   
   

\bibitem{Jin06}  Jin, B., h-Index: An evaluation indicator proposed by scientist.  {\sl Science Focus, 1} (1), 8-9  (2006).  
 

   \bibitem{Jin07} Jin, B., The AR-index: complementing the h-index.  {\sl ISSI Newsletter, 3} (1), 6 (2007).  

   \bibitem{JinRE07} Jin, B., Liang, L., Rousseau, R., \& Egghe, L., The R- and AR-indices: Complementing the h-index.  {\sl Chinese Science Bulletin, 52} (6), 855-863  (2007). 
   
\bibitem{1HBPhysA}   Bougrine, H., Subfield Effects on the Core of Coauthors,  Scientometrics, 98, 1047--1064   (2014). 

  \bibitem{1JMproceedingscore}    Mi\`skiewicz, J.,  Effects of Publications in Proceedings  on the Measure of the Core Size of Coauthors,  Physica A  392,   5119--5131 (2014). 
     
\bibitem{binarystars}  Ausloos, M.,  Binary Scientific Star Core Size of Coauthors,   Scientometrics,  1-21 (2014).
 
\bibitem{Iinamadrid}  Hellsten, I.,  Lambiotte, R.,  Scharnhorst, A.,   \&  Ausloos, M.,  Self-citations networks as traces of scientific careers. In Proceedings of the ISSI 2007, 11th International Conf. of the Intern. Society for Scientometrics and Informetrics, CSIC, Madrid, Spain, June 25-27, 2007. Ed. by D. Torres-Salinas; H. Moed, Vol. 1, pp. 361-367  (2007).

          \bibitem{egg1}  Egghe, L.,  \&  Rousseau, R. Introduction to Informetrics. Quantitative Methods in Library, Documentation and Information Science, Elsevier, Amsterdam  (1990). 
          
             \bibitem{WCS1.09.57}  Hadi, A. S.,  Rahmatullah Imon,  A. H. M.,  \&   Werner, M.,  Detection of outliers, {\it WIREs Comput Stat. 1}, 57-70 (2009).  
  
  \bibitem{WCS5.13.30}   Rojo, J.,  Heavy-tailed densities, {\it WIREs Comput Stat. 5}, 30-40  (2013). 
  
      \bibitem{8} Slone, R. M.   Coauthors contributions to major papers published in the AJR: Frequency of undeserved coauthorship.  {\sl American Journal of Radiology 167},  571-579 (1996). 
          
 \bibitem{Perssonetal04}    Persson,  O.,  Gl\" anzel,  W.,  \&  Danell,  R.  , Inflationary Bibliometric Values: The Role of. Scientific Collaboration and the Need for Relative Indicators in Evaluative Studies.    {\sl Scientometrics  60}, 421-432   (2004). 
  
          \bibitem{galam} Galam, S.,  Tailor based allocations for multiple authorship: a fractional gh-index.  {\sl Scientometrics 89}, 365-379  (2011).
          
            \bibitem{Egghebook05}     Egghe, L., Power Laws in the Information Production Process: Lotkaian Informetrics. Elsevier Academic Press
  (2005). 
           
\end{thebibliography}
\end{document}